\documentclass[showpacs,prd,twocolumn,aps,nofootinbib,superscriptaddress]{revtex4-1}

\usepackage{amsmath}
\usepackage{amssymb}
\usepackage{latexsym}
\usepackage{amsfonts}
\usepackage{epsfig}
\usepackage{psfrag}
\usepackage{graphicx}
\usepackage{wasysym}
\usepackage{ulem}
\usepackage{color}

\newcommand{\Eqref}[1]{Eq.~\eqref{#1}}

\renewcommand{\vec}[1]{\mathbf{#1}}

\begin{document}

\title{Quantum Reflection as a New Signature of Quantum Vacuum Nonlinearity}

\author{Holger Gies}
\author{Felix Karbstein}
\author{Nico Seegert}
\affiliation{Theoretisch-Physikalisches Institut, Abbe Center of Photonics,
Friedrich-Schiller-Universit\"at Jena, Max-Wien-Platz 1, D-07743 Jena, Germany}
\affiliation{Helmholtz-Institut Jena, Fr\"obelstieg 3, D-07743 Jena, Germany}

\begin{abstract}
 We show that photons subject to a spatially inhomogeneous
 electromagnetic field can experience quantum reflection.  Based on
 this observation, we propose quantum reflection as a novel means to
 probe the nonlinearity of the quantum vacuum in the presence of
 strong electromagnetic fields.
\end{abstract}

\date{May 10, 2013}


\maketitle

\section{Introduction}

The fundamental interaction of light and matter is described by
quantum electrodynamics (QED).  In contrast to classical
electrodynamics, the QED vacuum is no longer characterized by the
complete absence of any field excitations, but can rather be
considered as permeated by virtual photons and particle-antiparticle
fluctuations.  As these virtual fluctuations can couple to real
electromagnetic fields or matter, they have the potential to affect
the propagation and interactions of real fields and particles and can be probed accordingly.

The most prominent examples are the Casimir effect \cite{Casimir:dh},
revealing fluctuation-induced matter--matter interactions,
and nonlinear self-interactions of the electromagnetic field
  induced by electron-positron vacuum fluctuations
  \cite{Heisenberg:1935qt,Weisskopf,Schwinger:1951nm}. The latter
  example gives rise to a variety of nonlinear vacuum phenomena such
  as light-by-light scattering \cite{Euler:1935zz,Karplus:1950zz},
  vacuum magnetic birefringence
  \cite{Toll:1952,Baier,BialynickaBirula:1970vy}, photon splitting
  \cite{Adler:1971wn}, and even spontaneous vacuum decay in terms of
  Schwinger pair-production in electric fields
  \cite{Sauter:1931zz,Heisenberg:1935qt,Schwinger:1951nm}; for recent
  reviews, see
  \cite{Dittrich:2000zu,Marklund:2008gj,Dunne:2008kc,DiPiazza:2011tq}.
Whereas the Casimir effect has already been confirmed experimentally
\cite{Lamoreaux:1996wh,Mohideen:1998iz,Bressi:2002fr}, the pure
  electromagnetic nonlinearity of the quantum vacuum though subject to
  high-energy experiments \cite{Akhmadaliev:1998zz,Akhmadaliev:2001ik}
  has not been directly verified on macroscopic scales so
  far. Promising routes aim at vacuum magnetic birefringence
  measurements such as the PVLAS \cite{Cantatore:2008zz}, BMV \cite{Berceau:2011zz}
  experiments, or proposed set-ups on the basis of high-intensity
  lasers \cite{Heinzl:2006xc}.

In this paper our focus is on optical signatures, because modern
optical facilities allow for a huge photon number for probing, while
photon detection is possible even on the single-photon level. As
quantum vacuum nonlinearities can effectively be viewed as conferring
medium-like properties to the vacuum, a natural route is to search for
interference effects as suggested in \cite{King:2013am,Tommasini:2010fb,Hatsagortsyan:2011}.
By contrast, in the present work
we emphasize the viewpoint that strong electromagnetic fields can
modify the quantum vacuum such that the nonlinearly responding vacuum
acts as an effective potential for
propagating probe photons.

A highly sensitive probe of the shape of potentials is above-barrier
reflection \cite{QR1}, also called quantum reflection, as -- in
contrast to classical physics -- the barrier need not be repulsive
\cite{QR2}. Quantum reflection of atoms off a surface typically at
grazing incidence is nowadays commonly used in surface science
\cite{QR3,QR4}, and has even been applied to quantitatively measure
the fluctuation induced Casimir-Polder force \cite{Druzhinina:2003}.

In the present work, we suggest the use of quantum reflection as a new
way to explore the fluctuation-induced nonlinearities of the quantum
vacuum in a pump-probe type experiment. Replacing the atoms by photons
(``probe'') and the surface by a magnetized quantum vacuum (``pump''),
we obtain a highly sensitive set-up. In particular, a classical
background in the form of specular reflections, as is typical for
atomic quantum reflection, is completely absent in our case. There is
simply no analogue of a classical repulsive potential independently of
the incident angle.  Especially in comparison to standard
birefringence set-ups, where the induced quantum-vacuum signature has
to be isolated from a large background, e.g., by means of high-purity
polarimetric techniques \cite{Cantatore:2008zz,Marx:2011}, our
proposal of quantum reflection inherently allows for a clear
separation between signal and background, facilitating the use of
single-photon detection techniques.

Whereas the standard nonlinear phenomena listed above exist in
spatially homogeneous fields, quantum reflection manifestly requires
the external field to feature a spatial inhomogeneity. Below, we
discuss in the main body of the paper, how quantum reflection is
encoded in the quantum Maxwell equation by means of the
fluctuation-induced two-point correlation function (photon
polarization tensor). The relation to the more conventional language
of above-barrier scattering in quantum mechanics is highlighted in the
Appendix.

Our paper is organized as follows: Section~\ref{seq:qref} explains the
scenario of quantum reflection in detail.  The determination of the
photon reflection rate requires insights into the photon polarization
tensor in the presence of spatially inhomogeneous electromagnetic
fields.  A strategy to obtain the relevant analytical insights is
outlined in Sect.~\ref{seq:Piinh}. Here we limit ourselves to purely
magnetic fields.  Section~\ref{seq:Ex+Res} is devoted to the
discussion of explicit examples and results. We end with conclusions
and an outlook in Sect.~\ref{seq:Con+Out}.

\section{Quantum Reflection} \label{seq:qref}

We analyze the scenario of quantum reflection within the effective theory of photon propagation in a (spatially inhomogeneous) external magnetic field $\vec{B}(x)$.

The effective theory for soft electromagnetic fields in the quantum vacuum is provided by the famous Heisenberg-Euler Lagrangian \cite{Heisenberg:1935qt} and its generalizations to inhomogeneous backgrounds (cf., e.g., \cite{Gusynin:1998bt,Dunne:2000up,Gies:2001zp}).
Its generalization for photon propagation at arbitrary frequencies is described by the following Lagrangian (cf., e.g., \cite{Dittrich:2000zu}),
\begin{equation}
\mathcal{L}[A]= -\frac{1}{4} F_{\mu\nu} F^{\mu\nu}\!
- \frac{1}{2}\!\int_{x'}\! A_\mu(x) \Pi^{\mu\nu}\!(x,x'|\vec{B}) A_\nu(x'),\label{eq:calL}
\end{equation}
where $\Pi^{\mu\nu}(x,x'|\vec{B})$ denotes the photon polarization tensor in the presence of the
external field, $F_{\mu \nu}$ the field strength tensor of the propagating
photon $A_\mu$, and $x$ a spatio-temporal four-vector.
We use the metric convention $g_{\mu\nu}={\rm diag}(-,+,+,+)$, such that the momentum four vector squared reads $k^2=\vec{k}^2-(k^0)^2$.
Moreover, $c=\hbar=1$. Our conventions for the Fourier transform are $\Pi^{\mu\nu}(x,x')=\int_k\int_{k'}{\rm e}^{-ikx}\,\Pi^{\mu\nu}(k,k')\,{\rm e}^{-ik'x'}$ and $A_\mu(x)=\int_{k}\,{\rm e}^{ixk}A_\mu(k)$.

In momentum space, the equation of motion (``quantum Maxwell
equation'') associated with \Eqref{eq:calL} reads
\begin{equation}
 \left(k^2 g^{\mu \nu} - k^\mu k^\nu \right)\!A_\nu (k) = -\!\int_{k'}\!\tilde\Pi^{\mu\nu}(-k,k'|\vec{B})A_\nu(k') \label{eq:EOM} ,
\end{equation}
where we introduced the symmetrized polarization tensor $\tilde\Pi^{\mu\nu}(k,k'|\vec{B})=[\Pi^{\mu\nu}(k,k'|\vec{B})+\Pi^{\nu\mu}(k',k|\vec{B})]/2$.

Equation~\eqref{eq:EOM} is well suited to study the phenomenon of quantum reflection.
The basic idea is to interpret the right-hand side of \Eqref{eq:EOM} as source term for the reflected photons.
In this sense, the photon field on the right-hand side of \Eqref{eq:EOM} corresponds to the incident photon field, while the one on the left-hand side describes outgoing photons.

Equation~\eqref{eq:EOM} is a tensor equation of rather complicated structure. Fortunately, it can be simplified substantially by imposing additional constraints:
First, we limit ourselves to inhomogeneities of the form $\vec{B}(x)=B(x)\hat{\vec{B}}$, such that the direction of the magnetic field is fixed and only its amplitude is varied.
This defines a global spatial reference direction $\hat{\vec{B}}$, with respect to which vectors can be decomposed into parallel and perpendicular components,
\begin{equation}
  k^\mu=k^\mu_\parallel + k^\mu_\perp\,,\quad k_\parallel^\mu=(k^0,\vec{k}_\parallel)\,,\quad k_\perp^\mu=(0,\vec{k}_\perp)\,,
\end{equation}
with $\vec{k}_\parallel=(\vec{k}\cdot\hat{\vec{B}})\hat{\vec{B}}$ and $\vec{k}_\perp=\vec{k}-\vec{k}_\parallel$.
In the same way tensors can be decomposed, e.g., $g^{\mu\nu}=g_\parallel^{\mu\nu}+g_\perp^{\mu\nu}$.
For photons with four momentum $k^\mu$, it is then convenient to
introduce the following projectors \cite{Dittrich:2000zu},
\begin{equation}
 P^{\mu\nu}_\parallel(k)=g^{\mu\nu}_\parallel-\frac{k_\parallel^\mu k_\parallel^\nu}{k_\parallel^2}\,,\quad P^{\mu\nu}_\perp(k)=g^{\mu\nu}_\perp-\frac{k_\perp^\mu k_\perp^\nu}{k_\perp^2}\,. \label{eq:Proj}
\end{equation}
As long as $\vec{k}\nparallel\vec{B}$, the projectors~\eqref{eq:Proj}
have an intuitive interpretation. They project onto photon modes
polarized parallel and perpendicularly to the plane spanned by
$\vec{k}$ and $\hat{\vec{B}}$. Together with a third projector
defined as follows,
\begin{equation}
 P_0^{\mu\nu}(k)=g^{\mu\nu}-\frac{k^\mu k^\nu}{k^2}-P^{\mu\nu}_\parallel(k)-P^{\mu\nu}_\perp(k)\,,
\end{equation}
$P^{\mu\nu}_\parallel(k)$ and $P^{\mu\nu}_\perp(k)$ span the
transverse subspace.  For $\vec{k}\parallel\vec{B}$ only one
externally set direction is left, and we encounter rotational
invariance about the magnetic field axis. Here, the modes $0$ and
$\perp$ can be continuously related to the two zero-field polarization
modes \cite{Karbstein:2011ja}.

Second, we use that the field inhomogeneity can only affect momentum
components pointing along the inhomogeneity,
i.e., those (anti)parallel to $\pmb{\nabla}B$, while translational
invariance holds for the perpendicular directions.  Correspondingly,
we can identify two situations where \Eqref{eq:EOM} turns out to be
particularly simple:
$(i)$ If the magnetic field vector and the direction of the
inhomogeneity are orthogonal to each other (or
$\vec{k}_\parallel=0$),
\begin{equation}
 \vec{k}_\parallel\cdot\pmb{\nabla}B=0 \quad\to\quad P_\parallel^{\mu\nu}(k')=P_\parallel^{\mu\nu}(k)\equiv P^{\mu\nu}_\parallel\,, \label{eq:(i)}
\end{equation}
\Eqref{eq:EOM} can be simplified straightforwardly for the $\parallel$ polarization mode.
Contracting \Eqref{eq:EOM} with the global projector $P_\parallel$ and introducing photons $A_{p,\mu}(k)=P_{p,\mu\nu}A^\nu(k)$  polarized in mode $p\in\{\parallel,\perp\}$, the equation of motion loses any nontrivial Lorentz index structure.
Dropping the trivial Lorentz indices of the photon fields, $A_{p,\mu}(k)\to A_{p}(k)$, we obtain the scalar equation
\begin{equation}
 k^2A_{\parallel} (k) = -\int_{k'}\tilde\Pi_\parallel(-k,k'|B)A_{\parallel}(k') \label{eq:EOM(i)} \,.
\end{equation}
To arrive at \Eqref{eq:EOM(i)}, we also used the above reasoning for the photon polarization tensor, which for $B=const.$ is of the following structure \cite{Dittrich:2000zu} 
\begin{equation}
\Pi^{\mu\nu}(k,k'|\vec{B})\big|_{B=const.}=\delta(k+k')\!\!\!\sum_{p=\parallel,\perp,0}\!\!\!\Pi_p(k|B)P^{\mu\nu}_p\,, \label{eq:Pidec}
\end{equation}
with scalar coefficients $\Pi_p(k|B)$, carrying the entire field strength dependence.

$(ii)$ If the perpendicular component of the photon wave vector and the direction of the inhomogeneity are orthogonal to each other (or $\vec{k}_\perp=0$),
an analogous simplification holds for the $\perp$ polarization mode. The corresponding equations follow from Eqs.~\eqref{eq:(i)} and \eqref{eq:EOM(i)} by replacing $\parallel\to\perp$.
We limit ourselves to the discussion of the special cases $(i)$ and $(ii)$, since other configurations are more subtle as the
inhomogeneity genuinely induces mixings between the different polarization modes.

\begin{figure}
\vspace*{0.5cm}
\includegraphics[width=0.45\textwidth]{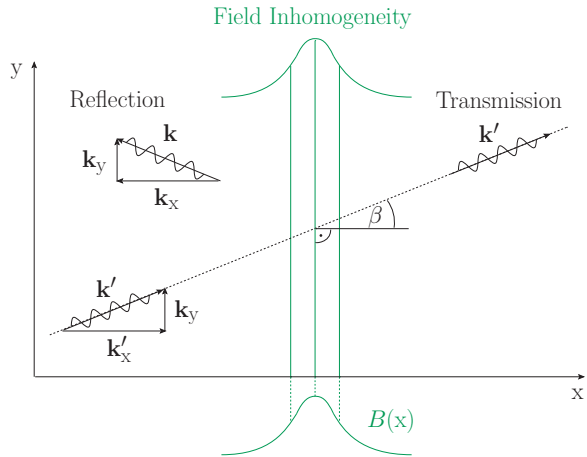} 
\caption{Schematic depiction of quantum reflection. The incident probe photons with wave vector ${\vec k}'$ propagate towards a one dimensional field inhomogeneity of amplitude $B({\rm x})$ which asymptotically falls off to zero for large values of $|{\rm x}|$.
The inhomogeneity is infinitely extended in the transversal directions. The angle between ${\vec k}'$ and $\vec{e}_{\rm x}$ is denoted by $\beta$.}
\label{fig:Qref}
\end{figure}

Without loss of generality we subsequently assume the field
inhomogeneity in ${\rm x}$ direction, i.e., $\pmb{\nabla}B\sim\vec{e}_{\rm x}$,
and limit ourselves to incident photons with wave vector
$\vec{k}'=k_{\rm x}'\vec{e}_{\rm x}+k_{\rm y}\vec{e}_{\rm y}$ (cf. Fig.~\ref{fig:Qref}).
This implies that the momentum component $k_{\rm y}$ is conserved, and thus also
inherited by the reflected photons, whose wave vector correspondingly
reads $\vec{k}=k_{\rm x}\vec{e}_{\rm x}+k_{\rm y}\vec{e}_{\rm y}$.  The scalar equations
derived for cases $(i)$ and $(ii)$ [cf.~\Eqref{eq:EOM(i)}] are then of
the following structure,
\begin{equation}
 \left(k_{\rm x}^2-\!(\omega^2\!-\!k_{\rm y}^2)\right)\!A_p(k)\! = \!-\!\int\!\frac{{\rm d}k_{\rm x}'}{2\pi}\,\tilde\Pi_p(\!-k_{\rm x},k'_{\rm x})A_p(k') \label{eq:EOMgen}\,, \!
\end{equation}
where $\omega$ is the photon frequency. To keep notations simple, we
have removed any reference to the magnetic field as well as the
conserved momentum components $k_{\rm y}$ and $\omega$ in the argument of
the photon polarization tensor.  Instead, its argument now only
includes the momentum components affected by the inhomogeneity, $k_{\rm x}$
and $k_{\rm x}'$.  Noteworthy, for the above reasoning it is not necessary
to explicitly specify the direction of $\hat{\bf B}$, which is however
implicitly constrained by demanding compatibility with either case
$(i)$ or $(ii)$.

Introducing partial Fourier transforms,
\begin{gather}
 A_p({\rm x})\equiv A_p({\rm x};k_{\rm y},\omega)=\int\frac{{\rm d} k_{\rm x}}{2\pi}\,{\rm e}^{ik_{\rm x}{\rm x}}A_p(k)\,, \\
 \tilde\Pi_p({\rm x},{\rm x}')=\int\frac{{\rm d} k_{\rm x}}{2\pi}\int\frac{{\rm d} k_{\rm x}'}{2\pi}\,{\rm e}^{-ik_{\rm x}{\rm x}}\,\tilde\Pi_p(k_{\rm x},k_{\rm x}'){\rm e}^{-ik_{\rm x}'{\rm x}'},
\end{gather}
\Eqref{eq:EOMgen} can alternatively be written as
\begin{equation}
 \left(\partial_{\rm x}^2+\tilde\omega^2\right)A_p({\rm x}) = \int{\rm d}x'\,\tilde\Pi_p({\rm x},{\rm x}')A_p({\rm x}') \label{eq:EOMx} \,,
\end{equation}
with $\tilde\omega^2\equiv\omega^2-k_{\rm y}^2$. 
This representation is particularly suited for the study of quantum reflection, as it directly allows for an intuitive physical approach to tackle the phenomenon in position space.
Here our focus is on a `localized' inhomogeneity $B({\rm x})$ of typical width $w$ which falls off to zero sufficiently fast for large values of $|{\rm x}|$.

We moreover assume all reflected photons to be detected independently of the particular value of $k_{\rm y}$. Formally, this amounts to detectors spanning the entire ${\rm y}$ axis.
However, photon reflection only takes place in a limited interval of typical diameter $w$ where $B({\rm x})$ deviates from zero.
For this assumption to hold with regard to an actual experimental realization we therefore just have to demand detector sizes compatible with the length scale of the inhomogeneity.
An inhomogeneity of width $w$ requires a detector size of the order of $w_{\rm eff}=2w\tan\beta$ in ${\rm y}$ direction (cf. Fig.~\ref{fig:Qref}).

In order to handle this theoretically, we assume the probe photons, emitted by a photon source located at ${\rm x}=-L$ with $B(-L)=0$, to be purely right-moving, i.e.,
$A^{\rm in}_p({\rm x}')=a(\tilde\omega){\rm e}^{i\tilde\omega{\rm x}'}$.
Here we have made use of the light-cone condition $k_{\rm x}^2-\tilde\omega^2=0$, neglecting subleading light cone deformations inside the magnetic field.
We then look for outgoing left-moving ($\hat{=}$ reflected) photons at detector positions ${\rm x}''<-L$.

On the level of \Eqref{eq:EOMx}, we identify the photon field $A_p({\rm x}')$ on its right-hand side with the incident photon field $A^{\rm in}_p({\rm x}')$.
In turn, \Eqref{eq:EOMx} may be interpreted as the equation of motion for the outgoing photons in the presence of the photon source $j_{p}({\rm x})$,
\begin{equation}
 \left(\partial_{\rm x}^2+\tilde\omega^2\right)A^{\rm out}_p({\rm x}) = j_{p}({\rm x}) \label{eq:EOMAref} \,,
\end{equation}
with $j_p({\rm x})= \int_{-L}^\infty{\rm d}{\rm x}'\,\tilde\Pi_p({\rm x},{\rm x}')A^{\rm in}_p({\rm x}')$.
The propagation of the outgoing photons arising from the interaction
is assumed to be well-described by the free Green's function, satisfying
\begin{equation}
 \left(\partial_{\rm x}^2+\tilde\omega^2\right)G({\rm x},{\rm x}')=\delta({\rm x}-{\rm x}') \label{eq:Green}\,,
\end{equation}
the solution of which reads
\begin{equation}
 G({\rm x},{\rm x}')=-\frac{i}{2\tilde\omega}
          \begin{cases}
           {\rm e}^{+i\tilde\omega({\rm x}-{\rm x}')}\quad\text{for}\quad {\rm x}-{\rm x}'>0\,, \\
           {\rm e}^{-i\tilde\omega({\rm x}-{\rm x}')}\quad\text{for}\quad {\rm x}-{\rm x}'<0\,.
          \end{cases}
\end{equation}
Asymptotically, the upper line (together with the incoming photons) is associated with the transmitted photons, whereas the lower line corresponds to the reflected photons $A_p^{\rm ref}$ which
we straightforwardly obtain from
\begin{equation}
 A_p^{\rm ref}({\rm x}''<-L)=-\frac{i}{2\tilde\omega}\int_{-L}^\infty{\rm d}{\rm x}\,j_p({\rm x})\,{\rm e}^{-i\tilde\omega({\rm x}''-{\rm x})}\,.
\end{equation}

The photon reflection coefficient can be defined via the ratio of reflected to incident photons.
Inserting the explicit expressions for the photon fields it can be represented in a particularly concise form,
\begin{equation}
 R_p=\lim_{L\to\infty}\left|\frac{A_p^{\rm ref}(x''<-L)}{A^{\rm in}_p(x')}\right|^2 
  =\left|\frac{\tilde\Pi_p(\tilde\omega,\tilde\omega)}{2\tilde\omega}\right|^2. \label{eq:R}
\end{equation}
The formal limit $L\to\infty$ is well-justified for `localized' inhomogeneities.
Thus, the reflection coefficient can be expressed entirely in terms of the photon polarization tensor in momentum space $\tilde\Pi(k_{\rm x},k_{\rm x}')$,
evaluated at $k_{\rm x}=\tilde\omega$, $k_{\rm x}'=\tilde\omega$ and made dimensionless by dividing by the momentum transfer $k_{\rm x}+k_{\rm x}'=2\tilde\omega$.
In particular note that this result is compatible with the light-cone condition for both incident and reflected photons, $k_{\rm x}^2-\tilde\omega^2=k_{\rm x}'^2-\tilde\omega^2=0$.

Finally, we recall \Eqref{eq:(i)} and emphasize again that the result~\eqref{eq:R} for $p=\parallel$ is associated with the particular alignment $(i)$, while that for $p=\perp$ belongs to situation $(ii)$.

\section{Photon polarization in spatially inhomogeneous fields} \label{seq:Piinh}

In a next step, we aim at analytical insights into the photon polarization tensor for spatially inhomogeneous magnetic fields, being intimately related to the photon reflection coefficient via \Eqref{eq:R}.
Whereas the corresponding expression in the presence of a homogeneous (electro)magnetic field is explicitly known at one-loop accuracy
\cite{Dittrich:2000zu,BatShab,Urrutia:1977xb,Schubert:2000yt}, no analytical results are available for generic, spatially inhomogeneous fields. Numerical insights are available from worldline Monte Carlo simulations \cite{Gies:2011he,Karbstein:2011ja}. 

Here our strategy is to focus on a situation sufficiently close to the constant field limit,
such that the photon polarization tensor for homogeneous magnetic fields can serve as a starting point for our considerations.
This should be true for field configurations that may be locally approximated by a constant:
In position space, the photon polarization tensor probes distances of the order of the virtual particles' Compton wavelength $\lambda_c= 1/m$, where 
$m$ corresponds to the electron mass in QED.
For inhomogeneities with a typical scale of variation $w$ 
much larger than the Compton wavelength of the virtual particles,
$w\gg \lambda_c$, using the constant-field expressions locally
is well justified.  For
electrons, $\lambda_c\approx2\cdot10^{-6}{\rm
  eV}^{-1}\approx3.9\cdot10^{-13}{\rm m}$.

We aim at the momentum space representation of the photon
polarization tensor locally accounting for a one-dimensional field
inhomogeneity $B({\rm x})$. This involves several steps, which can
be represented schematically as follows,
\begin{multline}
 \Pi^{\mu\nu}(k_{\rm x}')\,(2\pi)\,\delta(k_{\rm x}+k_{\rm x}')\ \xrightarrow{\rm F.T.}\ \Pi^{\mu\nu}({\rm x}-{\rm x}') \\
\xrightarrow{B \to B({\rm x})}\ \Pi^{\mu\nu}({\rm x},{\rm x}')\ \xrightarrow{{\rm F.T.}^{-1}}\ \Pi^{\mu\nu}(k_{\rm x},k_{\rm x}')\,, \label{eq:proc}
\end{multline}
where ${\rm F.T.}^{(-1)}$ refers to an (inverse) Fourier transform, and we have again only focused on components affected nontrivially by the inhomogeneity.

The photon polarization tensor for $B=const.$ in momentum space is explicitly known at one-loop accuracy \cite{Tsai:1974fa,Tsai:1975iz}: $\Pi^{\mu\nu}(k_{\rm x})$.
Translational invariance dictates its Fourier transform to depend on the relative coordinate ${\rm x}'-{\rm x}$ only: $\Pi^{\mu\nu}({\rm x}-{\rm x}')$.
Switching to a spatially inhomogeneous field by substituting $B\to B({\rm x})$ this invariance is broken: $\Pi^{\mu\nu}({\rm x},{\rm x}')$.
Transforming back to momentum space, the resulting polarization tensor $\Pi^{\mu\nu}(k_{\rm x},k_{\rm x}')$ mediates between two distinct momenta.

For $B=const.$ the photon polarization tensor has the following infinite series expansion,
\begin{equation}
 \Pi^{\mu\nu}(k_{\rm x}')=\sum_{n=0}^\infty \Pi^{\mu\nu}_{(2n)}(k_{\rm x}')(eB)^{2n}\,, \label{eq:SumPihom}
\end{equation}
which -- as a consequence of Furry's theorem -- is in terms of even powers of $eB$ only.
The expansion coefficients $\Pi^{\mu\nu}_{(2n)}(k_{\rm x}')$, with $n\in\mathbb{N}_0$ can be read off from a Taylor expansion of the photon polarization tensor for $B=const.$ to the desired order. 
In principle, each term $\Pi^{\mu\nu}_{(2n)}(k_{\rm x}')$ can be given in closed form. As these terms are rather lengthy -- even for $n=1$ --, we do not provide explicit expressions here.
Implementing the steps outlined in \Eqref{eq:proc} for \Eqref{eq:SumPihom},
we obtain
\begin{equation}
 \Pi^{\mu\nu}(k_{\rm x},k_{\rm x}')=\sum_{n=0}^\infty \Pi^{\mu\nu}_{(2n)}(k_{\rm x}')
 \int{\rm d}{\rm x}\,{\rm e}^{i(k_{\rm x}'+k_{\rm x}){\rm x}}[eB({\rm x})]^{2n}, \label{eq:SumPiinh0}
\end{equation}
and upon symmetrization [cf. \Eqref{eq:EOM}],
\begin{multline}
 \tilde\Pi^{\mu\nu}(k_{\rm x},k_{\rm x}')=\sum_{n=0}^\infty \frac{1}{2}\Bigl[\Pi^{\mu\nu}_{(2n)}(k_{\rm x}')+ \Pi^{\mu\nu}_{(2n)}(k_{\rm x}) \Bigr] \\
\times\int{\rm d}{\rm x}\,{\rm e}^{i(k_{\rm x}'+k_{\rm x}){\rm x}}[eB({\rm x})]^{2n}\,. \label{eq:SumPiinh}
\end{multline}
As the photon polarization tensor at zero field vanishes on the light cone, $\Pi^{\mu\nu}_{(0)}\big|_{k^2=0}=0$, the leading contribution to the photon reflection coefficient in the perturbative limit, $\frac{eB}{m^2}\ll1$, thus reads [cf. \Eqref{eq:R}]
\begin{equation}
R_p
=\Biggl|\frac{c_p}{\pi}\tilde\omega\int{\rm d}{\rm x}\,{\rm e}^{i2\tilde\omega{\rm x}}\left(\frac{eB({\rm x})}{m^2}\right)^{2}
\Biggr|^2 + {\cal O}\left((\tfrac{eB}{m^2})^{6}\right), \label{eq:res}
\end{equation}
with
\begin{equation}
\left\{\!
 \begin{array}{c}
c_\parallel\\
c_\perp
 \end{array}
\!\right\}
=
\frac{\alpha}{180}\left[\sin^2\theta+\sin^2\theta'\right] \left(\frac{\omega}{\tilde\omega}\right)^2
\left\{\!
 \begin{array}{c}
7\\
4
 \end{array}
\!\right\}, \label{eq:cs}
\end{equation}
where the angles $\theta'=\varangle(\vec{k}',\vec{B})$ and $\theta=\varangle(\vec{k},\vec{B})$
($\theta',\theta\in0\ldots\pi$) can differ for the kinematics of case {\it (ii)}, even though they
are still related by momentum conservation.  An alternative derivation
of the reflection coefficient~\eqref{eq:res} \'{a} la quantum
mechanics is given in the Appendix~\ref{sec:appendix}.  As expected,
the structure of \Eqref{eq:res} is similar to quantum mechanical
scattering in the Born approximation.

Recall that the first component, $R_\parallel$, corresponds to the result associated with situation $(i)$, while the second component, $R_\perp$, provides the result for $(ii)$.
We emphasize that \Eqref{eq:res} is valid for arbitrary profiles $B({\rm x})$ of a `localized' field inhomogeneity of width $w\gg\lambda_c$.
It is completely capable to deal with the field strengths attainable in present and near future laser facilities, and will form the basis of our subsequent considerations. 
For completeness, we note that the photon polarization tensor for $B=const.$ can be cast into the form \cite{Dittrich:2000zu}
\begin{equation}
 \Pi^{\mu\nu}(k_{\rm x}')\sim N^{\mu\nu}(k_{\rm x}'){\rm e}^{-i\frac{f(k_{\rm x}')}{eB}}\,, \label{eq:schemPi}
\end{equation}
i.e., its entire $B$ dependence occurs in the phase.
Employing \Eqref{eq:proc} in \Eqref{eq:schemPi}, we find
\begin{equation}
 \Pi^{\mu\nu}(k_{\rm x},k_{\rm x}')\sim N^{\mu\nu}(k_{\rm x}')\int{\rm d}x\,{\rm e}^{i(k_{\rm x}+k_{\rm x}')x}\,{\rm e}^{-i\frac{f(k'_{\rm x})}{eB({\rm x})}}. \label{eq:schemPi2}
\end{equation}
Thus, for inhomogeneities of the form $B({\rm x})=\frac{B}{1+({\rm x}/w)^2}$ the
integration in \Eqref{eq:schemPi2} is of Gaussian type and can be
performed explicitly.  Correspondingly, the full one-loop photon
polarization tensor in the presence of such an inhomogeneity can
eventually be written in terms of a double parameter integral of
similar complexity as for $B=const.$ This opens up the possibility to
study also manifestly nonperturbative effects in the presence of a
field inhomogeneity. This is, however, outside the scope of the present paper and
subject of an ongoing study.

\section{Results and Discussion} \label{seq:Ex+Res}

It is now straightforward and instructive to analytically determine the photon reflection coefficient for various forms of the field inhomogeneity $B({\rm x})$.
To keep notations compact, we subsequently only state the contribution due to the term written explicitly in \Eqref{eq:res} and omit any explicit reference to the neglected corrections, which are of ${\cal O}((\frac{eB}{m^2})^{6})$.
While, of course, a plethora of interesting field inhomogeneities is conceivable, here we exemplarily discuss only three generic shapes. 
Let us first consider two  elementary shapes, which do not exhibit any substructure and whose spatial form is solely characterized by a width parameter $w$:
A Lorentz profile is conventionally characterized by its full width at half maximum (FWHM).
For a magnetic field of this type,
\begin{equation}
 B({\rm x})=\frac{B}{1+\left(\frac{2{\rm x}}{w}\right)^2}\,, \label{eq:Blorentz}
\end{equation}
which decreases power-like for large values of $|\frac{{\rm x}}{w}|$, i.e., $\lim_{|x/w|\to\infty}B({\rm x})=(2{\rm x}/w)^{-2}$, \Eqref{eq:res} results in 
\begin{equation}
R_p=\biggl|\frac{c_p}{4}
\biggl(\frac{eB}{m^2}\biggr)^2\tilde\omega w(1+\tilde\omega w){\rm e}^{-\tilde\omega w}
\biggr|^2 , \label{eq:powerlike}
\end{equation}
which becomes maximal for $w=\frac{1+\sqrt{5}}{2\tilde\omega}\approx\frac{1.61}{\tilde\omega}$.
The reflection coefficient is exponentially suppressed with $\tilde\omega w$.
Conversely, for a field inhomogeneity of Gaussian type -- characterized by its full width at $1/{\rm e}$ of its maximum --
,
\begin{equation}
 B({\rm x})=B{\rm e}^{-(2{\rm x}/w)^2}\,, \label{eq:Bgauss}
\end{equation}
which falls off exponentially for large $|\frac{{\rm x}}{w}|$, we encounter an exponential suppression with $(\tilde\omega w)^2$,
\begin{equation}
 R_p=\biggl|\frac{1}{2}\frac{c_p}{\sqrt{2\pi}}\biggl(\frac{eB}{m^2}\biggr)^2\tilde\omega w\,{\rm e}^{-\frac{1}{8}(\tilde\omega w)^2}\biggr|^2 , \label{eq:RGauss}
\end{equation}
and find a maximum at $w=\frac{2}{\tilde\omega}$.

Next we turn to a more complicated field profile $B({\rm x})$, which besides its width $w$, is characterized by a modulation frequency $\omega_m$ (wavelength $\lambda_m$) and a phase $\varphi$.
As an example we consider the modulated Gaussian,
\begin{equation}
 B({\rm x})=B{\rm e}^{-(2{\rm x}/w)^2}\cos\left(\omega_m {\rm x}+\varphi\right)\,. \label{eq:Bmod}
\end{equation}
which results in the following photon reflection coefficient
\begin{multline}
 R_p=\biggl|\frac{1}{8}\frac{c_p}{\sqrt{2\pi}}\biggl(\frac{eB}{m^2}\biggr)^2\tilde\omega w \Bigl[2\,{\rm e}^{-\frac{1}{8}(\tilde\omega w)^2} \\
+\left({\rm e}^{-\frac{1}{8}w^2(\tilde\omega-\omega_m)^2-2i\varphi}
+{\rm e}^{-\frac{1}{8}w^2(\tilde\omega+\omega_m)^2+2i\varphi}\right)\Bigr]\biggr|^2 . \label{eq:modGauss}
\end{multline}
In the limit $\omega_m=0$, $\varphi=0$ this expression reduces to \Eqref{eq:RGauss}.
Most notably, for large values of $\{\tilde\omega w,\omega_m w\}\gg1$, but $\tilde\omega\simeq\omega_m$, \Eqref{eq:modGauss} becomes independent of $\varphi$, and is well approximated by
\begin{equation}
 R_p\approx\biggl|\frac{1}{8}\frac{c_p}{\sqrt{2\pi}}\biggl(\frac{eB}{m^2}\biggr)^2\tilde\omega w\, {\rm e}^{-\frac{1}{8}w^2(\tilde\omega-\omega_m)^2}\biggr|^2 , \label{eq:modGaussapprox}
\end{equation}
i.e., the exponential suppression of the reflection coefficient can be overcome by matching the (reduced) frequency of the probe photons $\tilde\omega$ with the modulation frequency, setting $\tilde\omega=\omega_m$.

To achieve a sizable reflection rate, the magnetic field strength, which enters the reflection coefficient as provided in \Eqref{eq:res} in the fourth power $\sim\left(\frac{eB}{m^2}\right)^4$, has to be large.
On a laboratory scale, field strengths of sufficient size are presently only attainable within the focal spots of high-intensity laser systems.
This suggests to probe the phenomenon of quantum reflection in an all optics {\it pump--probe} setup:
While the field inhomogeneity is generated in the focal spot of one high-intensity laser of wavelength $\lambda_{\rm pump}$, it is probed with another high-intensity laser (wavelength $\lambda_{\rm probe}$).
A purely magnetic field inhomogeneity could, e.g., be realized by superimposing two counter propagating laser beams.

For given laser parameters (wavelength $\lambda$ $\leftrightarrow$ photon energy $\omega=2\pi/\lambda$; pulse energy $\cal E$, and pulse duration $\tau$) the mean intensity $I={\cal E}/(\tau\sigma)$ in the focus,
and thus the mean field strength $B\approx\sqrt{2I}$, can be maximized by minimizing the focus cross-section area $\sigma=\pi(d/2)^2$, where $d$ is the beam diameter.
In generic high-intensity laser experiments $\sigma$ cannot be chosen at will, but is limited by the diffraction limit.
Assuming Gaussian beams, the effective focus area is conventionally defined to contain $86\%$ of the beam energy ($1/e^2$ criterion for the intensity).
The minimum value of the beam diameter in the focus, i.e., twice its {\it waist spot size}, is then given by $d=2f^\#\lambda$ \cite{Siegman}, 
with $f^\#$ the so-called $f$-number, defined as the ratio of the focal length and the diameter of the focusing aperture;
$f$-numbers as low as $f^\#=1$ can be realized experimentally.
Thus, within the focus of the {\it pump} laser field strengths of the order of
\begin{equation}
 B_{\rm pump}\approx\sqrt{0.86\,\frac{2}{\pi}\frac{{\cal E}_{\rm pump}}{\tau_{\rm pump}\lambda_{\rm pump}^2}}
\label{eq:Bpump}
\end{equation}
are attainable. 

Let us for the moment assume the effect of quantum reflection to be insensitive to the actual longitudinal profile of the pump laser pulse, such that we may approximate its longitudinal profile
by its envelope, and for $\tau_{\rm pump}\gg\lambda_{\rm pump}$ as roughly constant.
With these oversimplifying assumptions which will be critically examined below, we discuss two particular settings.

$(a)$ In the most straightforward experimental setting to imagine, the pump laser beam propagates along the ${\rm y}$ axis, while its transversal profile,
parametrized by the coordinate ${\rm x}$, evolves along the well-defined envelope of a Gaussian beam.
In its focus the transversal profile of the pump beam indeed matches a Gaussian field inhomogeneity~\eqref{eq:Bgauss}.
Correspondingly, the width $w$ of the field inhomogeneity can be identified with the focus diameter $d_{\rm pump}$, such that for $f^\#=1$, we have $w\approx2\lambda_{\rm pump}$.

Assuming that the diffraction spreading of the pump beam about its waist is sufficiently small, or equivalently, its Rayleigh range is large enough,
the beam diameter in the vicinity of the waist remains approximately constant and an experimental setting resembling Fig.~\ref{fig:Qref} is conceivable:
The {\it probe} beam hits the pump beam under an angle $\beta$ (cf. Fig.~\ref{fig:Qref}), and the reflected photons are detected with a photon counter placed accordingly.

Most noteworthy, this implies a setup inherent geometric separation of the reflected photons from both the photons of the pump laser and the transmitted part of the probe beam, such that the background is expected to be very low.
The clear geometric signal to background separation makes quantum reflection a particularly interesting candidate to probe the quantum vacuum nonlinearity.

$(b)$ Along the same lines, we can imagine a somewhat more involved experimental setting to induce a modulated inhomogeneity resembling \Eqref{eq:Bmod}:
Suppose we have two identical Gaussian beams with the above properties propagating -- within their confocal parameters -- parallel to each other along the ${\rm y}$ direction.
By means of a relative phase shift of $\lambda_{\rm pump}/2$, the two lasers can be adjusted such that the direction of the magnetic field in the focus of the first laser beam points exactly in the opposite direction as for the second one. 
De-tuning their beam axes by a distance of $\lambda_{\rm pump}$, their foci overlap and a modulated inhomogeneity of width $w\approx2\lambda_{\rm pump}$ and wavelength of modulation $\lambda_m\approx2\lambda_{\rm pump}$ is generated.

Let us emphasize, that there is no compelling reason to motivate that the longitudinal, and thus in particular the temporal profile of the laser pulse should be irrelevant for the effect of quantum reflection.
In particular, note that the time needed for probe photons to traverse the inhomogeneity, $t=cw\simeq 2c\lambda_{\rm pump}$, already corresponds to two temporal cycles of the pump, and thus
does not justify to approximate the inhomogeneity as stationary. The stationarity assumption rather holds for inhomogeneities of width $w\ll\lambda_{\rm pump}$, i.e., requires focusing beyond the diffraction limit and $f^\#\lesssim1$.
For a quantitative prediction of the photon reflection rates for realistic laser experiments,  also the temporal variation of the field inhomogeneity has to be accounted for.

As our study is the first to propose quantum reflection as a signature of the nonlinearity of the quantum vacuum, we will nevertheless insert some realistic laser parameters in the derived formulae.
The intention is to provide a first estimate of the magnitude of the effect considered here.

In order to maximize the effect, in \Eqref{eq:cs} we set $\theta=\theta'=\frac{\pi}{2}$, i.e., probe propagation direction and $B$ field are orthogonal, and adopt $w=2\lambda_{\rm pump}$ (cf. the discussion above).
For the other parameters we exemplarily adopt the design parameters of the two high-intensity laser systems to be available in Jena \cite{Jena}:
JETI~200 \cite{JETI200} ($\lambda_{\rm probe}=800{\rm nm}\approx4.06{\rm eV}^{-1}$,
${\cal E}_{\rm probe}=4{\rm J}\approx2.50\cdot10^{19}{\rm eV}$, $\tau_{\rm probe}=20{\rm fs}\approx30.4{\rm eV}^{-1}$)
and POLARIS \cite{POLARIS} ($\lambda_{\rm pump}=1030{\rm nm}\approx5.22{\rm eV}^{-1}$,
${\cal E}_{\rm pump}=150{\rm J}\approx9.36\cdot10^{20}{\rm eV}$, $\tau_{\rm pump}=150{\rm fs}\approx228{\rm eV}^{-1}$).
The magnetic field strength of the pump is obtained from \Eqref{eq:Bpump} and reads $eB_{\rm pump}/m^2=3.33\cdot 10^{-4}$.
The number of photons per pulse follows from $N={\cal E}/\omega$, implying $N_{\rm probe}=1.61\cdot10^{19}$, and $N_{\rm pump}=7.80\cdot10^{20}$.

Equations~\eqref{eq:powerlike}, \eqref{eq:RGauss} and \eqref{eq:modGauss} share the overall factor
\begin{equation}
\left\{\!\!
 \begin{array}{c}
c_\parallel\\
c_\perp
 \end{array}
\!\!\right\}\! \biggl(\frac{eB}{m^2}\biggr)^2\tilde\omega w
\,\to \, 
\frac{2\pi\alpha}{45\cos\beta}\left\{\!
 \begin{array}{c}
7\\
4
 \end{array}
\!\right\}\frac{\lambda_{\rm pump}}{\lambda_{\rm probe}}\biggl(\!\frac{eB_{\rm pump}}{m^2}\!\biggr)^2, \label{eq:vonhiernachda}
\end{equation}
encoding the full field strength dependence, but differ in the exponential terms. 
In the following discussion, our focus is on the field inhomogeneities discussed in (a) and (b), which are roughly compatible with the Gaussian beam scenario outlined above.
More specifically, we only adopt \Eqref{eq:RGauss} and the approximate expression for the modulated inhomogeneity, \Eqref{eq:modGaussapprox}, in this context now valid for $\lambda_{\rm probe}\simeq\lambda_m\cos\beta$.
Correspondingly, \Eqref{eq:modGaussapprox} can be written as
\begin{multline}
 R_p=\frac{\pi\alpha^2}{64800\cos^2\beta}
\left\{\!
 \begin{array}{c}
49\\
16
 \end{array}
\!\right\}
\biggr(\frac{\lambda_{\rm pump}}{\lambda_{\rm probe}}\biggl)^2\biggl(\!\frac{eB_{\rm pump}}{m^2}\!\biggr)^4 \\
 {\rm e}^{-(2\pi)^2\bigl(\frac{\lambda_{\rm pump}}{\lambda_{\rm probe}}\cos\beta-\frac{\lambda_{\rm pump}}{\lambda_m}\bigr)^2}. \label{eq:modapproxLaser}
\end{multline}
The result for the Gaussian inhomogeneity~\eqref{eq:Bgauss}, follows from \Eqref{eq:modapproxLaser} by multiplication with a factor of $16$ and sending $\lambda_m\to\infty$ [cf. Eqs.~\eqref{eq:RGauss} and \eqref{eq:modGaussapprox}].

To obtain the actual experimental observable, namely the number $N_p$ of reflected photons polarized in mode $p$, the reflection coefficient has to be multiplied with the number of incident probe photons.
This implies that the number of reflected photons {\it per shot} can be estimated as
\begin{equation}
 N_p= R_p f_{\rm int} N_{\rm probe} \,, \label{eq:NR}
\end{equation}
where we have introduced a factor $f_{\rm int}={\rm
  min}\{1,\frac{\tau_{\rm pump}}{\tau_{\rm probe}}\}$, providing a
first estimate of the fraction of the number of incident probe photons
interacting with the inhomogeneity.

For given laser parameters, the only free parameter in case of the Gaussian inhomogeneity~\eqref{eq:Bgauss} is the angle $\beta$.
The condition for \Eqref{eq:RGauss} to become maximum translates into $\cos\beta = \frac{1}{2\pi} \frac{\lambda_{\rm probe}}{\lambda_{\rm pump}}$.
Inserting the maximum condition in \Eqref{eq:modapproxLaser} with $\lambda_m\to\infty$ and multiplying with the factor of $16$, we obtain
\begin{equation}
 R_p=\frac{2\pi^3\alpha^2{\rm e}^{-1}}{2025}
\left\{\!
 \begin{array}{c}
49\\
16
 \end{array}
\!\right\}
\biggr(\frac{\lambda_{\rm pump}}{\lambda_{\rm probe}}\biggl)^4\biggl(\!\frac{eB_{\rm pump}}{m^2}\!\biggr)^4 , \label{eq:modapproxLaser0}
\end{equation}
and, for the explicit values of the laser systems given above, $\beta\approx82.9^\circ$ and
\begin{equation}
 R_p=
\left\{\!
 \begin{array}{c}
9.94\\
3.24
 \end{array}
\!\right\}\cdot10^{-19}\quad\to\quad 
N_p
\approx\left\{\!\!
 \begin{array}{c}
16.00\\
5.22
 \end{array}
\!\!\right\}. \label{eq:modapproxLaser1}
\end{equation}
For completeness, let us remark that analogous considerations for the Lorentz profile inhomogeneity~\eqref{eq:Blorentz}, plugging in the same parameters, yield angles and rates of the same order of magnitude.

Conversely, for the modulated Gaussian~\eqref{eq:modGauss} the modulation wavelength $\lambda_m$ provides an additional handle.
To overcome the exponential suppression, we aim at matching $\lambda_{\rm probe}=\lambda_m\cos\beta$.
Assuming a modulation with $\lambda_m=2\lambda_{\rm pump}$ as discussed in $(b)$, the matching condition implies $\cos\beta=\frac{1}{2}\frac{\lambda_{\rm probe}}{\lambda_{\rm pump}}$, such that \Eqref{eq:modapproxLaser} becomes
\begin{equation}
 R_p=\frac{\pi\alpha^2}{16200}
\left\{\!
 \begin{array}{c}
49\\
16
 \end{array}
\!\right\}
\biggr(\frac{\lambda_{\rm pump}}{\lambda_{\rm probe}}\biggl)^4\biggl(\!\frac{eB_{\rm pump}}{m^2}\!\biggr)^4.
\end{equation}
For the explicit laser parameters given above we obtain $\beta\approx67.2^\circ$, as well as
\begin{equation}
 R_p=
\left\{\!
 \begin{array}{c}
1.71\\
0.56
 \end{array}
\!\right\}\cdot10^{-20}\quad\to\quad 
 N_p
\approx\left\{\!\!
 \begin{array}{c}
0.28\\
0.09
 \end{array}
\!\!\right\}. \label{eq:modapproxLaser2}
\end{equation}
Note that the values encountered for the angles are rather large (cf. Fig.~\ref{fig:Qref}). Smaller angles are accessible with stronger modulations.

One might wonder why the explicit values given in \Eqref{eq:modapproxLaser2} are smaller than those in \Eqref{eq:modapproxLaser1} as the modulation was initially motivated as a means to overcome the exponential suppression.
Recall, however, that we have effectively managed to overcome the exponential suppression in both of our examples by also allowing for an adjustment of the angle parameter $\beta$ to maximize the respective reflection coefficient.
Correspondingly, the absolute size of the reflection coefficient and thus the number of reflected photons is governed by the numerical prefactors. These turn out to be favorable in the first example.

Finally, we emphasize again that the explicit values given in
Eqs.~\eqref{eq:modapproxLaser1} and \eqref{eq:modapproxLaser2} are
first estimates, as for the moment we have completely ignored the {\it
  longitudinal evolution} of the pump laser pulse.  They rather amount
to a first guess of the order of magnitude of the reflection
coefficients and the absolute numbers of quantum reflected photons per
shot. Based on these numbers, quantum reflection might be within reach 
with state of the art high-intensity laser systems.  Definitive
statements require 
to account for
the longitudinal variation -- and thus in particular the temporal
structure and evolution -- of field inhomogeneities.

\section{Conclusions and Outlook} \label{seq:Con+Out}

In this paper, we have studied quantum reflection as a new signature of the nonlinearity of the quantum vacuum in strong electromagnetic fields.
In contrast to the traditional signatures,
quantum reflection manifestly requires an inhomogeneous field configuration.

Limiting ourselves to a spatially inhomogeneous, but stationary
magnetic field, we have obtained first insights into this new
phenomenon.  We have devised a strategy to obtain analytical insights
for field inhomogeneities close enough to the constant field limit,
as to justify the slowly varying field approximation for the quantum correlations.  As the underlying approximation holds for
inhomogeneities whose typical scale of variation is much larger than
the Compton wavelength of the electron, many field inhomogeneities available in the laboratory can be dealt with within
this framework.

Looking for reflected photons in the field free region, we expect to
achieve a clear geometric signal to background separation, rendering
quantum reflection a particularly interesting candidate to probe the
quantum vacuum nonlinearity in strong laser fields.  Let us also
emphasize that the effect has a huge potential to be enhanced and
optimized, e.g., by modeling particularly suited field inhomogeneities
that maximize the reflection coefficient by exploiting constructive
interferences. Also ``two-color'' laser set-ups as will become
  available at Jena will be helpful to suppress background noise by
  suitable filtering.

First estimates of the magnitude of the effect for present day laser parameters are promising.
However, in order to allow for solid quantitative predictions of the effect for realistic laser experiments, we will eventually have to explicitly account for the temporal structure of the pump laser pulse.
This question is currently under investigation and will be addressed in a follow up study.

\section*{Acknowledgments}

We thank M.~C.~Kaluza, J.~Polz and M.~Zepf for interesting and enlightening
discussions. HG acknowledges support by the DFG under grants Gi
328/5-2 (Heisenberg program) and SFB-TR18.

\appendix

\section{Determination of the reflection coefficient \`a la quantum mechanics}\label{sec:appendix}

Let us briefly outline an alternative way to arrive at the
result~\eqref{eq:res}. For clarity and to avoid further complications
we stick to a magnetic field oriented orthogonal to the direction of
photon propagation, i.e., $\varangle(\vec{k},\vec{B})=\frac{\pi}{2}$,
and assume $\beta=\varangle(\vec{k},\vec{e}_{\rm x})=0$.  The basic idea is
to first derive the equations of motion for photons in the presence of
a weak homogeneous magnetic field, i.e., keeping terms up to ${\cal
  O}\left((\frac{eB}{m^2})^2\right)$, and to implement the transition
from $B=const.$ to a spatially inhomogeneous magnetic field $B({\rm x})$ on
this level only.

Employing the photon dispersion relation for weak electric fields, $k^2=0+{\cal O}\left((\frac{eB}{m^2})^2\right)$,
for photons polarized in mode $p$ the equations of motion in momentum space can be straightforwardly approximated as follows,
\begin{equation}
 \left(k^2+\Pi_p(k|B)\right)A_p(k)=0\,, \label{eq:A1}
\end{equation}
with
\begin{equation}
 \left\{
 \begin{array}{c}
 \Pi_\parallel\\ 
 \Pi_\perp
 \end{array}
\right\}
=-\frac{\alpha}{45\pi}
\left\{
\begin{array}{c}
7\\
4
\end{array}
\right\}\left(\frac{eB}{m^2}\right)^2\omega^2.
\end{equation}
Thus only two polarization components $p\in\{\parallel,\perp\}$ exhibit a nontrivial dependence on the external field amplitude (cf. also \cite{Dittrich:2000zu,Karbstein:2011ja}).

For these modes, a Fourier transform to position space results in the following one dimensional Schr\"odinger equation,
\begin{equation}
 \left[-\frac{d^2}{d{\rm x}^2}-\omega^2\left(1+2\frac{c_p}{\pi}\left(\frac{eB}{m^2}\right)^2\right)\right]A_p({\rm x};\omega)=0\,,
\end{equation}
with $c_\parallel=7\alpha/90$ and $c_\perp=4\alpha/90$ [cf. \Eqref{eq:cs}], and after the replacement $B\to B({\rm x})$,
\begin{equation}
 \left(-\frac{d^2}{d{\rm x}^2}+V({\rm x})\right)A_p({\rm x};\omega)=\omega^2 A_p({\rm x};\omega)\,, \label{eq:QM}
\end{equation}
where we introduced the spatially localized {\it potential}
\begin{equation}
 V({\rm x})=-2\frac{c_p}{\pi}\omega^2\left(\frac{eB({\rm x})}{m^2}\right)^2. \label{eq:potential}
\end{equation}

The quantum mechanical scattering problem as posed by \Eqref{eq:QM} can be conveniently solved in the transfer matrix formalism, discretizing the spatial coordinate as ${\rm x}_n=n\Delta {\rm x}$ with $n\in\mathbb{N}$, and correspondingly the potential as $V_n=V(n\Delta {\rm x})$,
such that the dispersion relation for the $n$th step reads $k_n=\sqrt{\omega^2-V_n}$.

In the determination of the transfer matrix for the infinitesimal step $\Delta {\rm x}$ from ${\rm x}_{n}$ to ${\rm x}_{n+1}$
 -- and analogously for the multiplication of the transfer matrices for subsequent steps -- 
we assume the ratio $\frac{\Delta k}{\Delta {\rm x}}$ as finite and keep only terms up to ${\cal O}(\Delta {\rm x})$.
Reverting to the continuum limit the components of the transfer matrix for macroscopic distances can eventually be written in terms of integrals. 
Assuming left and right moving contributions at ${\rm x}=-\infty$, but just a right-moving component at ${\rm x}=\infty$,
we can straightforwardly derive an expression for the {\it quantum mechanical} reflection coefficient,
\begin{equation}
 R_p=\left|\frac{\int_{-\infty}^{+\infty}{\rm d}{\rm x}\,{\rm e}^{i2{\rm x}k({\rm x})}\frac{k'({\rm x})}{2k({\rm x})}}{1+\int_{-\infty}^{+\infty}{\rm d}{\rm x}\left(\frac{k'({\rm x})}{2k({\rm x})}+i{\rm x}k'({\rm x})\right)}\right|^2, \label{eq:RQM}
\end{equation}
with $k({\rm x})=\sqrt{\omega^2-V({\rm x})}$ and $k'({\rm x})=\frac{d}{d{\rm x}}k({\rm x})$.
For the potential~\eqref{eq:potential}, \Eqref{eq:RQM} results in
\begin{equation}
 R_p=\left|\frac{c_p}{\pi}\omega\int{\rm d}x\,{\rm e}^{i2\omega {\rm x}}\left(\frac{eB({\rm x})}{m^2}\right)^2\right|^2 + {\cal O}\left(\left(\tfrac{eB}{m^2}\right)^{6}\right), \label{eq:resQM}
\end{equation}
which is fully compatible with \Eqref{eq:res}.
Of course, also \Eqref{eq:resQM} yields well-defined results only when either condition $(i)$ or $(ii)$ discussed in the main text [cf. \Eqref{eq:(i)}] is met.

\end{document}